\documentclass[letterpaper,twocolumn,10pt]{article}
\usepackage{usenix-2020-09}
\usepackage{cite}
\usepackage{amsmath,amssymb,amsfonts}
\usepackage{algorithmic}
\usepackage{graphicx}
\usepackage{textcomp}
\usepackage{subcaption}
\usepackage{makecell}
\usepackage{balance}
\usepackage{xcolor}
\usepackage{float}
\PassOptionsToPackage{usenames,dvipsnames}{xcolor}
\usepackage{tikz}
\usepackage{authblk}

\def\BibTeX{{\rm B\kern-.05em{\sc i\kern-.025em b}\kern-.08em
    T\kern-.1667em\lower.7ex\hbox{E}\kern-.125emX}}

\newcommand*\circled[1]{\tikz[baseline=(char.base)]{
            \node[shape=circle,draw,inner sep=2pt] (char) {#1};}}
\newcommand{\ignore}[1]{}

\begin{document}

\title{\Large \bf SoK: An Introspective Analysis of RPKI Security}

%\author[$\S\ddag$]{Donika Mirdita}
%\author[$\S\dag$]{Haya Schulmann}
%\author[$\S\ddag$]{Michael Waidner}
%\affil[$\ddag$]{Technische Universität Darmstadt\hspace{25pt}$\dag$Goethe-Universität Frankfurt\hspace{25pt}$\S$ATHENE}

\author{Donika Mirdita$^{\dagger\natural}$\qquad Haya Schulmann$^{\ddagger\natural}$\qquad Michael Waidner$^{\dagger\natural}$\vspace{2mm}\\
{\small
$^\dagger$ Technische Universität Darmstadt \qquad $^\ddagger$ Goethe-Universität Frankfurt \qquad
$^\natural$ ATHENE} \\
}

\maketitle

\begin{abstract}
The Resource Public Key Infrastructure (RPKI) is the main mechanism to protect inter-domain routing with BGP from prefix hijacks. It has already been widely deployed by large providers and the adoption rate is getting to a critical point. Almost half of all the global prefixes are now covered by RPKI and measurements show that 27\% of networks are already using RPKI to validate BGP announcements.\\
\indent Over the past 10 years, there has been much research effort in RPKI, analyzing different facets of the protocol, such as software vulnerabilities, robustness of the infrastructure or the proliferation of RPKI validation. In this work we compile the first systemic overview of the vulnerabilities and misconfigurations in RPKI and quantify the security landscape of the global RPKI deployments based on our measurements and analysis. Our study discovers that 56\% of the global RPKI validators suffer from at least one documented vulnerability. We also do a systematization of knowledge for existing RPKI security research and complement the existing knowledge with novel measurements in which we discover new trends in availability of RPKI repositories, and their communication patterns with the RPKI validators. We weave together the results of existing research and our study, to provide a comprehensive tableau of vulnerabilities, their sources, and to derive future research paths necessary to prepare RPKI for full global deployment.
\end{abstract}
\section{Introduction}

The Border Gateway Protocol (BGP) is the primary routing protocol on the Internet. BGP facilitates the exchange of packets between global Autonomous Systems (ASes) by distributing prefix ownership announcements. BGP applies fine-tuned policies to determine the optimal path to a destination\cite{bellovin1989security}. However, BGP was designed with simplicity and efficiency in mind. The protocol lacks any security checks, exposing the networks to prefix hijack attacks. Routers, without adequate security, can accept bogus route announcements, thus allowing malicious entities to manipulate traffic flow. Such attacks have been exploited to blackhole, redirect or subvert network traffic \cite{china:telecom,ballani2007study,fb:out,u:tube,mitm:threat,indosat:hijack,turkey:hijack,vervier2015mind}. 
There have been several proposals for security mechanisms for BGP \cite{irr,huston2011securing}, but the most promising protocol so far has been the RPKI.  \\
\indent RPKI was first formalized in RFC6480, purposefully designed to be compatible with existing router software and hardware capabilities. RPKI's primary purpose is validating resource ownership via Route Origin Authorizations (ROAs), which are cryptographically signed files that bind network prefixes to their owner AS. RPKI objects are stored in hierarchically distributed repositories, and are periodically fetched and validated by RPKI relying party validators.  
The validated objects, called Validated ROA Payloads (VRPs) are cached and fed to BGP routers, who then update their routing policy accordingly. This design introduces neither new onerous hardware requirements on routers, nor bottlenecks in packet processing. 
Border routers do not need to perform cryptographic computations and only upload data to populate the RPKI parameters and perform Route Origin Validation (ROV), i.e., filter bogus BGP announcements.\\ 
\indent RPKI adoption rates show an increasingly positive trend, and its incorporation in global Internet backbones\cite{decix-rpki,aws-rpki} is a sign that RPKI is poised to become the primary security infrastructure to protect BGP traffic. Currently, 27\% of global networks enforce ROV \cite{hlavacek2023keep}. Meanwhile, 
47.7\% of global
IPv4 and 50.45\% of IPv6 prefixes are covered by ROAs\cite{nist}, thereby positioning RPKI as an imminently critical protocol for Internet security. \\
\indent Over the past 10 years, considerable research has been done to measure and evaluate RPKI deployments \cite{wahlisch2015ripki,kristoff2020measuring}, discover novel attack vectors \cite{mirdita2022poster,van2022rpkiller,usenix-stalloris-21}, and analyze the RPKI standard for scalability issues and the RPKI software suite for zero-days \cite{shrishak2021privacy,hlavacek2023beyond,mirdita2023cure}. Our analysis shows that many software vulnerabilities continue to be present in the RPKI ecosystem despite CVE issuance and attempted patches.\\
\indent In this work, we present the first comprehensive security analysis of the RPKI ecosystem. We review existing literature on RPKI security risks and deployment trends. We perform new measurements of the RPKI ecosystem and network communications, systemize the research knowledge into an overview of all known vulnerabilities, attack vectors, persistent implementation issues and map out these vulnerabilities to the global RPKI deployment. We determine that at least 56\% of all validators on the Internet are vulnerable to known attack vectors, such as Denial of Service (DoS), silent ROV downgrades, path traversals, cache poisonings etc. Major RPKI software providers are affected, including Routinator which makes up 70.5\% of the validators' market share. We observe that despite the criticality of RPKI, developers and users are not sufficiently security conscious. Some validator developers react slowly to bug disclosures, while others suffer from a passive user-base that do not upgrade their software. \\
\indent Our research includes novel measurements, which detects new persistent issues in RPKI deployments: insufficient object and repository management. RPKI communications are often unstable, leading to unforeseen errors and dropped connections, which results in loss of ROAs from the standpoint of routers. Errors range from DNS issues, to unstable connections, to malformed objects and misconfigured webservers. Our data suggests that timeout and network errors are responsible for considerable delays during the validation process. Additionally, we observe inefficient resource management from the repository operators, which often lead to temporary blackouts of resources due to mismanaged certification and object formatting. These endemic errors present a considerable obstacle to full RPKI deployment and a risk to Internet stability at full RPKI deployment. \\
\indent {\bf Contributions.} We perform a novel longitudinal study of the vulnerabilities and misconfigurations in RPKI deployments, and quantify their pervasiveness on the Internet. We evaluate RPKI component communication channels, assess the quality of RPKI repository management, the rates and nature of persistent errors during RPKI validation, and extract insights on current trends. We  use our analysis to identify threats to the RPKI ecosystem. We complement our study with a systematization of the complete body of knowledge on RPKI security and RPKI attack vectors. Finally, we highlight the next steps for the community to improve RPKI and prepare it for the global routing stage.\\
\indent {\bf Organization.} In Section \ref{sec:background} we provide a background on the RPKI protocol. Section \ref{sok} is a standardization of RPKI security literature and Section \ref{sec:method} illustrates our testbed and new measurement choices. Section  \ref{sec:threat} systematizes attack vectors against RPKI and quantifies their spread on live deployments. In Section \ref{sec:errors} we report the results of our longitudinal measurements on repository management and RPKI network communication errors. In Section \ref{sec:eval} we analyze a major factor that continuously introduces issues in the RPKI: human negligence. In Section \ref{fw} we provide a summary of future research paths to protect and improve RPKI. Finally, in Section \ref{sec:conclusions}, we summarize our observations and derive final conclusions on the status of RPKI and next steps to follow.
\section{Technical Overview of RPKI}\label{sec:background}

RPKI consists of hierarchically distributed repositories called Publication Points (PPs) and Relying Party (RP) validator software that fetches, processes, and validates RPKI objects. RPs are responsible for processing RPKI objects and ultimately feeding routers via RPKI-To-Router (RTR) protocol with the necessary data to inform their routing choices. 
The RPKI protocol was designed to be flexible and serve not just ROAs, but a range of BGP security helper files such as BGPsec certificates[RFC8209], ASPA \cite{aspa} and other, potentially new RPKI objects.  \\
\indent RPKI relies on Regional Internet Registries (RIRs) to provide a chain of trust for all globally issued ROAs. RIRs give out Trust Anchor Locators (TALs) [RFC6490], which are files that contain a pointer towards the location of the RIR RPKI root certificate. RIR RPKI certificates contain a special header extension, where the location of the RIR RPKI repository is hardcoded. From that point onward the RPs can sequentially discover all global PPs and download their objects. \\
\indent {\bf Publication Points} are the building blocks of the RPKI distributed database. A PP serves a unique subset of global ROAs, with each PP serving the objects pertaining to one or more Certificate Authorities (CAs). CAs are the logical units of RPKI, each representing a resource owning entity. Each CA contains a certificate that hardcodes its resources. When a CA delegates ownership of resources to another entity, it stores a certificate for the new CA, which hardcodes the location and set of resources this new CA now owns. A child CA can be hosted either in the same PP as its parent or in an external one. 
RPKI can be enabled either in hosted or delegated mode. In hosted RPKI, the resource owner delegates ROA management to an external entity e.g. a RIR. In delegated RPKI, the resource owners create their own CA, and manage their own ROAs. These entities must provide to an RPKI service provider, e.g. a RIR, their CA certificate so their resources become discoverable.\\
\indent {\bf Relying Party} is the software component that periodically fetches, parses and cryptographically validates RPKI objects. RPs begin every validation round with the RIRs and process every discoverable PP along the way. After validation, RPs compile the Verified ROA Payloads (VRPs), a list of ASN-Prefix tuples stored in ROAs. VRPs are fed to BGP routers via the RPKI-To-Router protocol. 
An {\it invalid} VRP will disqualify that path from being used to forward packets, a {\it valid} VRP will prioritize that path, but when a path does not have any VRP entries, it is classified as {\it not found} and it is provisionally treated the same as a path with a valid RPKI status.\\  
\indent {\bf RPKI Objects.} Currently, the primary use of RPKI is ROA distribution. However, the RPKI ecosystem requires additional auxiliary objects to ensure security and integrity. Key objects in the RPKI ecosystem are manifests, certificates, and certificate revocation lists. Certificate revocation lists allow administrators to selectively revoke certificates. The manifest [RFC9286] is a periodically re-issued, short-lived file, which is responsible for repository content integrity. It contains a list of objects and their corresponding SHA-256 hash. 
Each repository requires a manifest. RPKI uses X.509 certificates [RFC3779] as attestation that a repository has the right to manage select Internet resources. 
RPKI objects are cryptographically signed by their CA's private key, and for every file changes, signatures are regenerated. Further, RPKI aims to broaden its offerings by incorporating additional security protocols, such as BGPsec and Autonomous System Provider Authorizations. RPKI is expected to become a framework that supports most standardized routing protection protocols, but right now, it only supports prefix origin validation.\\ 
\indent {\bf Chain of Trust} is a fundamental feature of RPKI. Trust begins with the root of the RPKI ecosystem, the 5 RIRs and their CA certificates. RIRs provide their TALs to all global users, so the moment an operator loads a TAL file, they are implicitly trusting this root and every signature their CAs issue. All objects in the repository are signed by CA private keys, including the certificate of child delegations. The presence of the RIR parent signature legitimizes the child CA resource ownership claims. As long as there is an uninterrupted chain of signed certificates that lead to the root RIR CA, we have a chain of trust for all CA resource claims in the RPKI tree. \\
\indent {\bf Connection Protocols.} RPs periodically fetch RPKI objects from PPs using one of two standardized protocols: RPKI Repository Delta Protocol (RRDP) and rsync. rsync was the protocol of choice during RPKI's initial deployment. However, it soon became inefficient due to its heavy processing requirements, susceptibility to DoS attacks, and lack of libraries\cite{arin}. RRDP[RFC8182] is an HTTPS-based protocol tailored to RPKI requirements. It is designed to avoid rsync pitfalls and introduce more security to RPKI component communications. RRDP allows RPs to perform full or partial content synchronization, thus minimizing network traffic and optimizing repository convergence. For this purpose, 3 file types were introduced in the RPKI ecosystem: the notification, delta and snapshot files. The notification file is the gateway to the repository content and the first file an RP encounters while accessing a new PP. The notification file contains the URI and hash for snapshot and deltas. The snapshot is the exhaustive list of all objects in the repository; the delta contains repository incremental changes. 
If an RP is querying a PP for the first time, it will by default do a full snapshot download and save the session\_id and serial\_id. During the follow-up validations, the RP performs a delta update if the session\_id remains unchanged and the RP can bootstrap all the delta increments from its repository cache without errors. If delta fails, the RP falls back to snapshot download. \\
\indent {\bf Validation Process}. Validation begins with the 5 RIR TALs. Each TAL contains a pointer to the CA certificate of the RIR's RPKI infrastructure. These certificates contain an extension, which hardcodes the repository's URI. 
RPs download the notification file, which contains the URI for the deltas and snapshot. Then, it evaluates which mode is most appropriate and downloads the listed objects. Next, RPs parse and validate the manifest. Only files in the manifest are then further validated. Object validation requires multiple evaluation steps per file and auxiliaries[RFC8360]. RPs validate cryptographic signatures to ensure repository integrity. \\
\indent {\bf RPKI-To-Router (RTR)} is a custom protocol that delivers cryptographically validated payloads to routers. This protocol was designed to fit within the constraints of modern routing platforms with minimal memory requirements. Routers cannot handle the extra load of cryptographic validation of RPKI data, so they receive easily digestible results computed by RPs. Routers establish a connection with the RP and continuously sync their cache with the RP's cache. The RTR standard encourages the use of protocols that provide authentication and integrity guarantees, e.g., IPsec. RTR also supports unprotected TCP connections with the caveat that both RP and router must be located in the same trusted network.

\begin{table}[H]
\scriptsize
\renewcommand{\arraystretch}{0.6}
    \centering
    \begin{tabular}{c|c|c|c}
         \textbf{Category} & \textbf{References} & \textbf{Area} & \textbf{Year}  \\ \hline
        \makecell{ROV\\Measurements} & \makecell{\cite{gilad2016we}\\\cite{reuter2018towards}\\ \cite{hlavacek2018practical}\cite{rodday2021revisiting}\\ \cite{apnic}\cite{cloudflare} \\ \cite{rovista}\cite{hlavacek2023keep}} & \makecell{passive measurements \\ active measurements \\ traceroute technique \\  active measurements \\ traceroute \& active measurements } & \makecell{2016 \\ 2018 \\ 2018/21 \\ 2021 \\ 2023}\\ \hline 
        \makecell{Attack\\Vectors} & \makecell{\cite{cooper2013risk} \\ \cite{van2022rpkiller}\cite{usenix-stalloris-21}\\ \cite{mirdita2022poster} \cite{DBLP:conf/ccs/HlavacekJMSW22} \\ \cite{mirdita2023cure}}& \makecell{censorship \\ ROV downgrade\\  \\ fuzzing RP} & \makecell{2013 \\ 2022/23\\ \\ 2023}  \\ \hline
       \makecell{Defense\\Mechanisms} & \makecell{\cite{shrishak2021privacy} \\ \cite{morillo2021rov++} \\ \cite{hlavacek2023beyond} \\ \cite{drr}} & \makecell{multi-party computation \\ ROV extension \\ validation re-ordering algorithm \\ distributed computing} &  \makecell{2021 \\ 2021 \\ 2023 \\ 2024} \\ \hline
        \makecell{Deployment} & \makecell{\cite{wahlisch2015ripki}\\\cite{testart2020filter}\\\cite{livadariu2024tale}\cite{zeng2024improving}} & \makecell{ROV web coverage \\ ISP IXP deployment strategy \\ global deployment strategy} & \makecell{2015 \\ 2020 \\ 2024}\\ \hline
        \makecell{Surveys} & \makecell{\cite{kristoff2020measuring} \\ \cite{rodday2023resource}} & \makecell{RP survey \\ literature survey} & \makecell{2020 \\ 2023} \\ \hline
    \end{tabular}
    \vspace{-5pt}
    \caption{A summary of RPKI security literature.}
    \label{tab:reslit}
\end{table}

\section{Analysis of RPKI Security Literature}\label{sok}

RPKI is a core routing security protocol and there is ample research on various security-related facets.
In this section, we deconstruct the current body of RPKI research into areas most relevant to the security and deployment of the RPKI ecosystem, see Table \ref{tab:reslit}. We determine 5 relevant research directions and provide a chronological review of the most important works in those areas. We also explain how this SoK enriches and extends the current state of knowledge.

\subsection{ROV Measurements} ROV measurements provide us with deployment milestones for RPKI. Comprehensive RPKI protection can only be ensured if ROV is widely enforced, and therefore considerable research was invested in measuring ROV deployment rates.
\\
\indent The first attempt to measure ROV \cite{gilad2016we} dates back to 2016 and passively measures a lower bound of non-adopting ASes that propagate RPKI invalid BGP paths. These measurements painted a bleak picture. At most 3\% of the 100 top Internet Service Providers (ISPs) enforced ROV. However, such passive measurements are prone to false positives, e.g., due to traffic engineering. Further research refined the methodology and introduced active measurement techniques to bypass existing limitations. \cite{reuter2018towards} complements passive measurements with active controlled experiments for the first time. The authors utilize the PEERING\cite{peering} testbed to announce multiple prefixes with different RPKI status. They inferred that an AS enforces ROV, if it correctly discriminates between both RPKI statuses. This technique had the downside of only classifying an AS as ROV enforcing if it peered with the researchers' testbed or provided a Looking Glass.  \\
 \indent \cite{testart2020filter} measured ROV deployment by looking at ASes that share their routes with BGP collectors and filter for the presence of RPKI invalid paths. Their measurements show at most 11\% of ASes enforce ROV. Then, \cite{cloudflare} introduced a crowd-sourcing approach by having volunteers manually log ROV enforcing ASes. However, this approach is limited by the participants' rigorosity in actively updating the list. In \cite{hlavacek2018practical}, the authors designed a new approach for ROV measurements, where they leverage traceroutes and TCP handshakes to detect ROV enforcement for on-path ASes. The authors use RIPE Atlas \cite{atlas} to launch their measurements, and probe websites from the 1.25M-top Alexa webservers \cite{alexa}. This technique was further refined by \cite{rodday2021revisiting}, which introduced filters that limit false positives, relaxed strong testbed connectivity assumptions, and differentiated between partial and full ROV enforcement. \\
 \indent \cite{apnic} goes one step further and enriches the set of ROV detection approaches by using multiple active measurements: a short-term experiment to measure AS sensitivity to ROA validity oscillations, and a conventional measurement filtering for ASes propagating invalid BGP announcements by issuing announcements with conflicting ROV status. The results showed a similar ROV enforcement rate to \cite{testart2020filter}, with at most 10\% of observed hosts residing inside ROV enforcing networks. In 2023, RoVista \cite{rovista} combined passive monitoring of RPKI invalid prefix propagation with active measurements to find that 12.3\% of analyzed ASes fully enforce ROV. However, RoVista based its active measurements on the assumption of hosts using IP-ID, a counter which is no longer widely used on the Internet according to recent work \cite{pearce2017augur,DBLP:conf/dsn/ShulmanZ21,dai2021smap}. Finally, \cite{hlavacek2023keep} extended ROV measurement techniques by utilizing globally distributed vantage points to send BGP announcements with competing ROV validity. They also utilize those vantage points to collect traceroutes to their controlled destinations. To reduce false positives, the authors introduced a metric which distinguishes ASes where path changes according to the ROV status of the destination prefix take place. This work finds that at least 27\% of global ASes enforce ROV. \\
\indent Existing ROV deployment research has supplied various measurement methodologies, active and passive, on the control- and data-plane. However, the collective downside of all existing approaches is the limited pool of ASes each paper evaluates. ROV measurements are usually limited to ASes peering with the author's testbed, providing a looking glass or ending up on the traceroute path, i.e., limited by the vantage point availability. More work is necessary in designing up-to-date methodologies that can measure ROV enforcement for the majority of active ASes on the Internet.

\subsection{Attack Vectors} Recent research highlights RPKI's vulnerability to zero-days and protocol design exploits. \cite{van2022rpkiller} is one of the first works to analyze RP software for vulnerabilities. Results showed many RP implementations were vulnerable to various attacks ranging from known gzip bombs via HTTP compression, XML billion laughs attack, and path traversals, to RPKI-specific issues like infinite processing subtrees, and coding errors leading to crashes due to malformed files. \cite{mirdita2022poster} reports two zero-days affecting the most popular RP implementations, that at the time affected 84.9\% of all global RPs with negligible cost. The zero-days were programming errors that crashed the software if objects were slightly malformed or if the size of the CA subtree crossed an arbitrary threshold. \\
\indent Then \cite{usenix-stalloris-21} introduced a novel attack mechanism to stealthily downgrade RPKI protection by exploiting vulnerabilities in the RPKI protocol. To instantiate this attack, an adversary needs to map the RPKI ecosystem: find where PPs and RPs are located on the Internet. This is a low-cost low-burden endeavor. The adversary only needs a PP in delegated mode, something any resource owning entity can set up, and an RP to extract PP information. With these two vantage points, anyone can map RPKI deployments via passive monitoring of RPKI traffic and collect query frequency, agent type and whereabouts of RPs, thus narrowing down RPs located in ASes of interest. Additional passive investigation using HTTP redirects, can uncover an RP target's DNS resolvers, thus forcing DNS requests to PP nameservers to be out-filtered by rate-limiting the servers with spoofed packets. While being part of the RPKI tree, an attacker can use connection slow down mechanisms, and unlimited PP delegations, to cause uninterrupted delay during a single validation round. Adversaries can time the attack with waiting out the manifest expiration of the victim repositories.
47\% of global PPs and 100\% of global RPs were vulnerable to this attack. \\
\indent Further, \cite{DBLP:conf/ccs/HlavacekJMSW22} discovered that 56.7\% of ASes have vulnerable DNS components that lead to new ways to disable RPKI. The community attempted to issue patches for the stalling attacks, but \cite{hlavacek2023beyond} showed that stalling is still viable in multiple forms for all RPs. In light of continuous discovery of crash inducing bugs in RP software, researchers \cite{mirdita2023cure} developed CURE, a dedicated generalistic and language agnostic black box RP fuzzer, in order to systemically analyze all RP implementations. It discovered 18 new vulnerabilities. Additionally, RPKI is also vulnerable to resource subversion and manipulation attacks, as described in \cite{cooper2013risk}. Resource delegation and revocation in RPKI can be done unilaterally by any resource owner high up in the hierarchy, without any checks, justification or verification. Researchers have shown that RPKI is vulnerable to targeted resource delegation manipulations that can affect arbitrary child or grandchild PPs. This exposes the protocol to censorship, since any PP including RIRs are subject to local legal enforcement. \\

\subsection{Defense Mechanisms} Current trends show an increased focus on research towards developing RPKI defenses. Researchers predominantly focus on improving the protocol by introducing security checks, subroutines to enforce object integrity, protocol redesigns to address censorship risks, and improve router-side ROV enforcement to maximize RPKI coverage for the Internet. \\
\indent \cite{shrishak2021privacy} is one of the first works tackling censorship in RPKI by proposing the use of multi-party computation protocols such as threshold signatures to protect the RPKI ecosystem from arbitrary object changes by compromised repositories, without a majority vote. This work was followed by \cite{drr}, who proposed a comprehensive partial-redesign of the RPKI software suite by replacing repositories with a federation of servers and monitors that utilize a leader-base byzantine fault tolerant protocol to reach a consensus on object changes. \\ 
\indent Other work focused on solving RP vulnerability against stalling attacks. \cite{hlavacek2023beyond} proposed improving the validation algorithm with a subroutine that calculates the moving average of download times for each PP in order to discern between malicious and benign PPs. This proposal was submitted as a solution for the ever-present stalling attacks. \\
\indent Finally, \cite{morillo2021rov++} designed ROV++, an extension to classic ROV, which requires no overhaul of the RPKI protocol itself and can be applied by routers with little to no changes on their RPKI interface. The purpose of this extension is to maximize RPKI protection throughout the Internet despite low deployment rates. Evaluation results showed increased protection from sub-/prefix-hijacking attacks in spite of low ROV adoption.\\
\indent Research in defense mechanisms is vital for improving RPKI robustness. However, it is important for vendors to incorporate these improvements, and, for security solutions to offer backwards compatibility with the existing software environment. Some proposals such as \cite{hlavacek2023beyond} and \cite{morillo2021rov++} are easy to incorporate in RPKI but those present only part of the solution. More work is necessary to fully patch the RPKI protocol and ensure continuous robustness of RPKI software deployments.

\subsection{Deployment Status} RPKI deployment faces an uphill battle in terms of global enforcement and coverage. Research from 2015\cite{wahlisch2015ripki} showed that the most popular websites had no protection and Content Distribution Networks (CDNs) ignored RPKI ROV. However, around that time, ISPs started to slowly deploy RPKI on their infrastructure. Not until 2019, did ROA coverage  reach $\sim$12.58\% of global IPv4 prefixes\cite{nist}. Then, \cite{testart2020filter} reports a growing number of ISPs and transit providers deploying ROV filtering. The authors estimate $\sim$10\% of providers in 2020 enforce ROV. Subsequent measurements in 2023\cite{hlavacek2023keep} increase the estimate to 27\%. Researchers\cite{livadariu2024tale} then analyzed how RPKI interoperates with other Internet Exchange Points (IXPs) technologies like Remotely Triggered Black Hole (RTBH), a DDoS mitigation technique. Analysis showed that ROV can conflict with RTBH, e.g., RTBH announcements sometime go counter to ROV statuses, leading to routers dropping the more specific RTBH announcements. Ideally, all routing security frameworks should be compatible instead of sabotaging each other's intended behavior. More research is required on the level of ISPs and IXPs to measure the compatibility of RPKI and other routing security mechanisms. \\
\indent While global deployment of RPKI ROV is necessary, waiting for all ASes to start enforcing can take an exorbitant amount of time given current deployment estimates. Recent research \cite{zeng2024improving} tries to solve the ROV coverage conundrum using evolutionary game theory. RPKI's deployment problem is a circular one: operators do not want to deploy it because it is still not widely used, but if they do not, RPKI remains a niche used by a minority. To solve this problem, the authors propose using BGP communities to introduce hijack notifications from ASes with RPKI ROV coverage, to neighbors that do not employ it. This approach increases global RPKI coverage, while warming up timid ASes to the benefits of employing their own RPKI infrastructure. 

\subsection{Standardization of Knowledge} 

Despite being a core security protocol, there have been few attempts in summarizing, analyzing and unpacking the wealth of RPKI research knowledge on surveys and papers. There are in fact only two such contributions.  \\
\indent On the one hand, we have \cite{kristoff2020measuring}, who performed the first collection of traces of RPKI RPs on the Internet to analyze their synchronisation patterns, and observed access protocol usage trends to identify mismatches between the protocol standard and actual software behavior. This work provided several interesting insights, such as the fact that operators are not inclined to change default parameters of software, leading to the overwhelming majority of RPs exhibiting the default refresh intervals. Additionally, at least 20\% of RPs did not obey the standard and would not fall back on rsync when RRDP fails. This work also highlights the presence of connection errors between RP and PP, but it does not delve into details on the topic of errors, failures, vulnerabilities in RPs, and their causes. On the other hand, there is \cite{rodday2023resource} which is the first comprehensive RPKI literature survey, discussing all facets of currently published RPKI research and standards. This work provides a full chronological overview of RPKI development and research, and insights on future development paths. \\
\indent In contrast to existing global surveys, our work focuses on the collection of indicators of vulnerability and the practical quantification of their presence in real-world RPKI deployments. We also analyze for the first time the nature of errors in the connections between RPs and PPs. We run longitudinal measurements and highlight the existence of patterns in the nature of re-occurring errors. We show for the first time a set of error types that downgrade RPKI performance globally, and highlight both the cause and solutions. Furthermore, we also contextualize the consequences of such errors and vulnerabilities in RPKI efficiency and resilience. We primarily focus on surveying papers that describe security coverage and vulnerabilities in RPKI, and run new measurements based on those insights with our customized testbed. \\
\indent We setup a measurements testbed and collect longitudinal data that allows us to discover new issues affecting live RPKI ecosystems on the Internet, and provide new insights on future work by the community to protect RPKI deployments. Based on our observations, we propose future work to resolve the obstacles towards RPKI deployment and to improve performance. To summarize the various attack vectors against RPKI software and versions, we compile the first comprehensive collection of all known CVEs against popular RP software implementations that are currently deployed on the Internet (in Table \ref{tab:rp-cve}). We use the MITRE and NIST CVE databases, and the security advisory pages of the RPs when available, to collect this data. We exclude from this table all RPKI-related CVEs that do not apply to RP or PP software. Based on the vulnerability distributions in Table \ref{tab:rp-cve}, and the data from our testbed, we quantify the number of RP instances on the Internet vulnerable to known attack vectors, and qualify the effects of these vulnerabilities on BGP routing security. \\
\indent We use our testbed to perform new measurements that allow us to offer the first analysis of the RPKI network connection patterns. Our measurements illustrate chronic issues in PP backend management, which cause unnecessary delays and disconnections, leading to RPs missing out on collecting the full set of global ROAs. We highlight, that RPKI requires future work not only on its protocol and software, but also on educating, supporting and improving usability for end users. We illustrate for the first time, how human error disproportionately affects RPKI security by introducing critical old vulnerabilities on the majority of deployments, even though solutions for most of these issues already exist.

\begin{figure}[t!]
  \centering
  \includegraphics[width=0.4\textwidth]{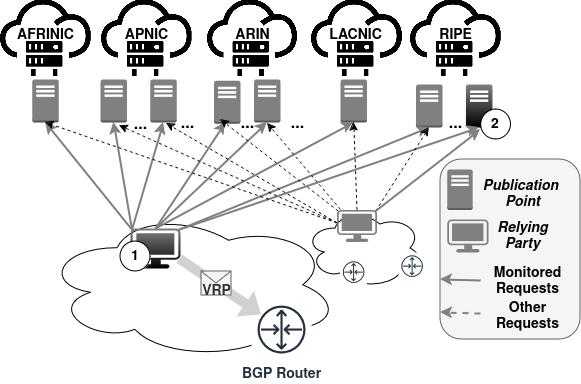}
  \vspace{-5pt}
  \caption{RPKI Testbed.}
  \label{fig:setuo}
\end{figure}

\section{Measurement Testbed}\label{sec:method}

In this work, we leverage existing research as indicators of problems and we run additional measurements to quantify the presence of vulnerabilities in the live RPKI ecosystem. We investigate new aspects of the RPKI ecosystem by conducting new complementary measurements to create a full global security tableau of RPKI. While \cite{kristoff2020measuring} mentions errors in the communication channel, the authors do not go into details on the nature of errors and their causes. We decide to run longitudinal measurements on the RPKI network as a mere passive participant. We run passive measurements to analyze the RPKI network as-is; we do not want to influence content and behavior. In order to discern and analyze security issues and deployment bottlenecks from the perspective of an RPKI operator, we aim to observe the RPKI ecosystem natively the same way as any RPKI user experiences it.\\
\indent To perform our measurements, we set up two monitoring instances on both sides of the RPKI system: we set up a monitoring RP and a monitoring PP. Figure \ref{fig:setuo} illustrates the RPKI infrastructure as a whole and highlights our testbed setup. Entities with circled numbers alongside them represent our monitoring infrastructure. We collect all communications exchanged between our monitors and the RPKI network. The collected connections are depicted in full undotted lines. \\
\indent \circled{1} is our RP instance, which we install on a well-connected network, and let it run for 2 years. We choose Routinator as the RP implementation of choice due its robustness, log verbosity, but also due to its popularity as the most widely used implementation in the world. We aim to observe the RPKI network as the vast majority of RPKI users experience it. The RP collects and stores information on all outgoing connections, as well as failures in establishing those connections, and data processing errors. Our instance uses the default Routinator refresh interval of 10 minutes. Through our RP instance, we observe the state, network speed, and content variability of all globally accessible PPs. \\
\indent We purchase network resources from RIPE Internet Registry and operate a LIR account on its platform. We use a VM to install the Krill publication point software \cite{krill0.9.3}, which allows us to create our own certificate authority and repository. We then register our Krill instance in our LIR portal as a PP in {\it delegated RPKI} mode. We host and serve our objects with our own instance, and RIPE only hosts a certificate pointing to our PP. This setup allows us to monitor and log all incoming RP requests from every RP instance in the world querying the RIPE TAL and moving down to our client. \circled{2} shows the position of our PP monitor inside the RPKI ecosystem. \\
\indent Our testbed is located in a network with a Tier-2 upstream provider. Our ISP provides us with a dedicated fiber link with speeds up to 1GBit/s. It peers with one of the largest IXPs in Europe as the direct transit provider connecting us directly to its routeserver and peers. Our data collection is complete and non-discriminating. We collect the full verbose logs our monitoring instances can generate. 
\section{RPKI Threat Taxonomy}\label{sec:threat}

In this section, we provide a systematization of all global RPKI threat vectors as discovered over the years. Our research establishes that vulnerabilities are consistently widespread across current RP deployments on the Internet.\\
\indent Currently, active RPKI vulnerabilities range from DoS, cache poisonings, remote code executions, buffer overflows, and path traversals, to complex compound attacks that lead to silent perpetual downgrade of ROV protection. The former type of vulnerability is straightforward to fix: these are well-known attacks and software patches have already been issued for most of them. The latter is more pernicious, as it exploits the logic of the protocol's validation process, for which there is no safeguard other than continuous monitoring of PP and RP behavior. In this study we observe lax security awareness by RPKI operators. On the one hand, we see developers who don't patch disclosed bugs in a timely way, on the other hand, operators unwilling to upgrade their software despite security risks.
In Table \ref{tab:rp-cve}, we find a compilation of all RPKI CVEs for currently used RP and PP software on the Internet.

\begin{table*}[h!]
\renewcommand{\arraystretch}{0.4}
    \centering
    \begin{tabular}{l|c|c|r|c|c}\hline
         \textbf{CVE} & \textbf{RPKI Component} & \textbf{Software} & \textbf{Versions} & \textbf{Attack Vector} & \textbf{Source} \\ \hline
         CVE-2023-39914 & rust object parsing & bcder & $\le$ 0.7.2 & DoS & \cite{mirdita2023cure}\\
         CVE-2022-39915 & Relying Party & Routinator & $\le$ 0.12.1 & DoS & \cite{mirdita2023cure}\\
         CVE-2022-39916 & Relying Party  &Routinator & 0.9.0 - 0.12.1 & \makecell{path traversal \\ RPKI poisoning} & \cite{mirdita2023cure} \\
         CVE-2023-0158 & Publication Point & Krill & $\le$ 0.12.0 & DoS & \cite{pp} \\
         CVE-2022-3616 & Relying Party &OctoRPKI & 1.3.0 - 1.4.4 &  DoS & \cite{mirdita2022poster}\\
         CVE-2022-3029 & Relying Party &Routinator & 0.9.0 - 0.11.2 & DoS & \cite{mirdita2022poster}\\
         CVE-2021-43174 & Relying Party &Routinator & 0.9.0 - 0.10.1 & out-of-memory crash & \cite{van2022rpkiller}\\
         CVE-2021-43173 & Relying Party &Routinator & $\le$ 0.10.1 & stalling attack & \cite{van2022rpkiller}\\ 
         CVE-2021-43172 & Relying Party & Routinator & $\le$ 0.10.1 & stalling attack & \cite{van2022rpkiller} \\
         CVE-2021-43114 & Relying Party &Fort & $<$ 1.5.2 & RTR DoS & \cite{van2022rpkiller}\\
         CVE-2021-41531 & Relying Party &Routinator& $<$ 0.9.0 & DoS & Job Snijders\\
         CVE-2021-3912 & Relying Party &OctoRPKI & $<$ 1.4.0 & out-of-memory crash & \cite{van2022rpkiller} \\ 
         CVE-2021-3911 & Relying Party &OctoRPKI & $<$ 1.3.0 & DoS & - \\ 
         CVE-2021-3910 & Relying Party &OctoRPKI & $<$ 1.3.0 & DoS & -\\
         CVE-2021-3909 & Relying Party &OctoRPKI & $<$ 1.3.0 & stalling attack & -\\
         CVE-2021-3908 & Relying Party &OctoRPKI & $<$ 1.3.0 & stalling attack & -\\
         CVE-2021-3907 & Relying Party &OctoRPKI & $<$ 1.4.3 & remote code execution & -\\
         CVE-2021-3761 & Relying Party &OctoRPKI & $<$ 1.3.0 & ROV DoS & Job Snijders \\ 
         CVE-2020-17366 & Relying Party &Routinator & $<$ 0.8.0 & DoS & Job Snijders\\
         CVE-2020-16162 & Relying Party & RIPE Validator & $\le$ 3.1-2020.07.06.14.28 & replay attack & Job Snijders \\
         CVE-2020-16163 & Relying Party & RIPE Validator & $<$ 3.1-2020.07.06.14.28 & bypass restrictions & Job Snijders \\
         CVE-2020-16164 & Relying Party & RIPE Validator & $\le$ 3.1-2020.07.06.14.28& bypass restrictions & Job Snijders \\ \hline
    \end{tabular}
    \vspace{-5pt}
    \caption{A chronicle of RPKI CVEs.}
    \vspace{-5pt}
    \label{tab:rp-cve}
\end{table*}

\subsection{Publication Point Attack Surface}

PPs are the resource repository of RPKI. PP blackouts affect ROV, and thus they require high availability and network stability. Currently, Krill \cite{krill0.9.3} is the only actively maintained open-source PP implementation. While many resource owners in delegated mode opt for Krill, it is not the universal choice. Major repository providers build their own customized PP implementations \cite{lacnic-rpki}. PPs are characterized by a small attack surface. Their computations are not influenced by external inputs; they generate, sign, and renew RPKI objects according to internal processing logic and custom configurations. Externally, their function is limited to that of a central repository that is queried by RPs via a simple interface. \\
\indent Despite their limited Internet facing interface, PPs are vulnerable to availability attacks that are not immediately visible, as this attack neither raises errors nor causes crashes. Worse, due to the common occurrence of network errors in the RPKI, lack of established connections can easily be mistaken for benign errors and go unnoticed even by monitors.

\subsection{Relying Party Deployments}
RPs are the middleman between PP repositories and BGP routers. They require high availability and must process and validate a wide range of files on account of several BGP security protocols. Due to their intricate processing logic, the most predominant choice between RPKI operators is to use one of the several open-source implementations. \\
\indent We measure the global distribution of RPs for December 2023.
Table \ref{tab:rp-dist} shows the global market share of the 5 main RPs in use. 99\% of all RP instances on the Internet belong to 1 of 5 open-source RP implementations. Routinator is the most popular RP with 70.5\% of the RP market share. rpki-client is the second most widely used RP, and OctoRPKI comes in 3rd place, followed by Fort in 4th place. We observe that 2.1\% of RPs run on RIPE NCC RPKI Validator, an RP that has been discontinued since 2021. The other 1.1\% is an aggregation of different querying requests with opaque agent headers.
\begin{table}[H]
\renewcommand{\arraystretch}{0.7}
    \centering
    \begin{tabular}{c|l|c|c}\hline
          & \textbf{Relying Party} & \textbf{Share} & \textbf{Status} \\ \hline
         1&Routinator & 70.5\% & in development \\
         2&rpki-client & 14.9\% & in development \\
         3&OctoRPKI & 5.7\% & in development \\
         4&Fort  & 5.6\% & in development \\
         5&RIPE Validator & 2.1\% & discontinued in 2021 \\
         6&others & 1.1\% & unknown \\\hline
    \end{tabular}
    \vspace{-5pt}
    \caption{Relying party deployments.}
    \label{tab:rp-dist}
\end{table}  

\subsection{Attacks Disabling Relying Party Validation}

\begin{figure}[H]
  \centering
  \vspace{-10pt}
  \begin{subfigure}[t]{0.49\columnwidth}
    \centering
    \includegraphics[width=\linewidth]{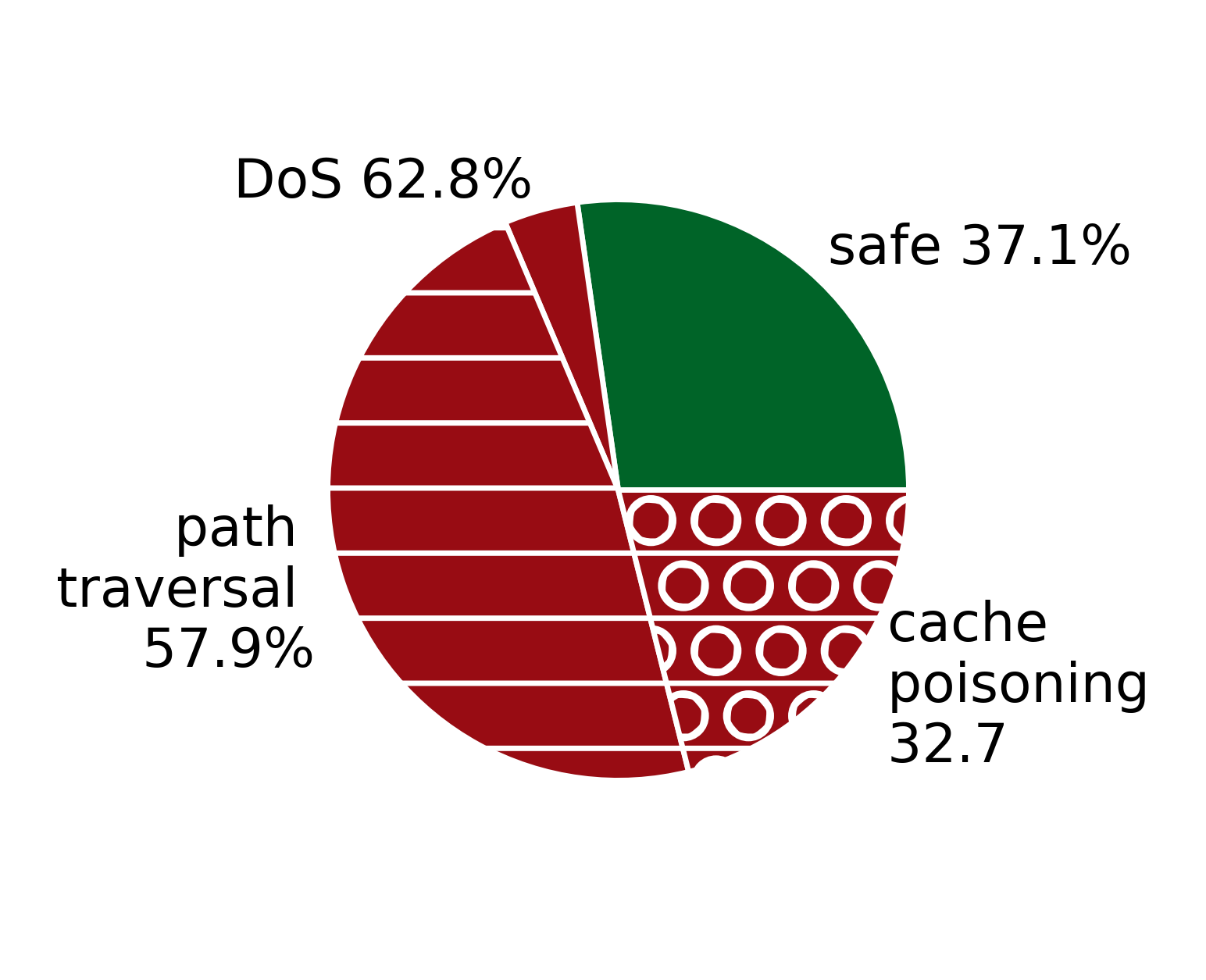}
    \vspace{-30pt}
    \caption{\footnotesize{Routinator}.}
    \label{fig:rout-vuln}
  \end{subfigure}
\begin{subfigure}[t]{0.49\columnwidth}
    \includegraphics[width=\linewidth]{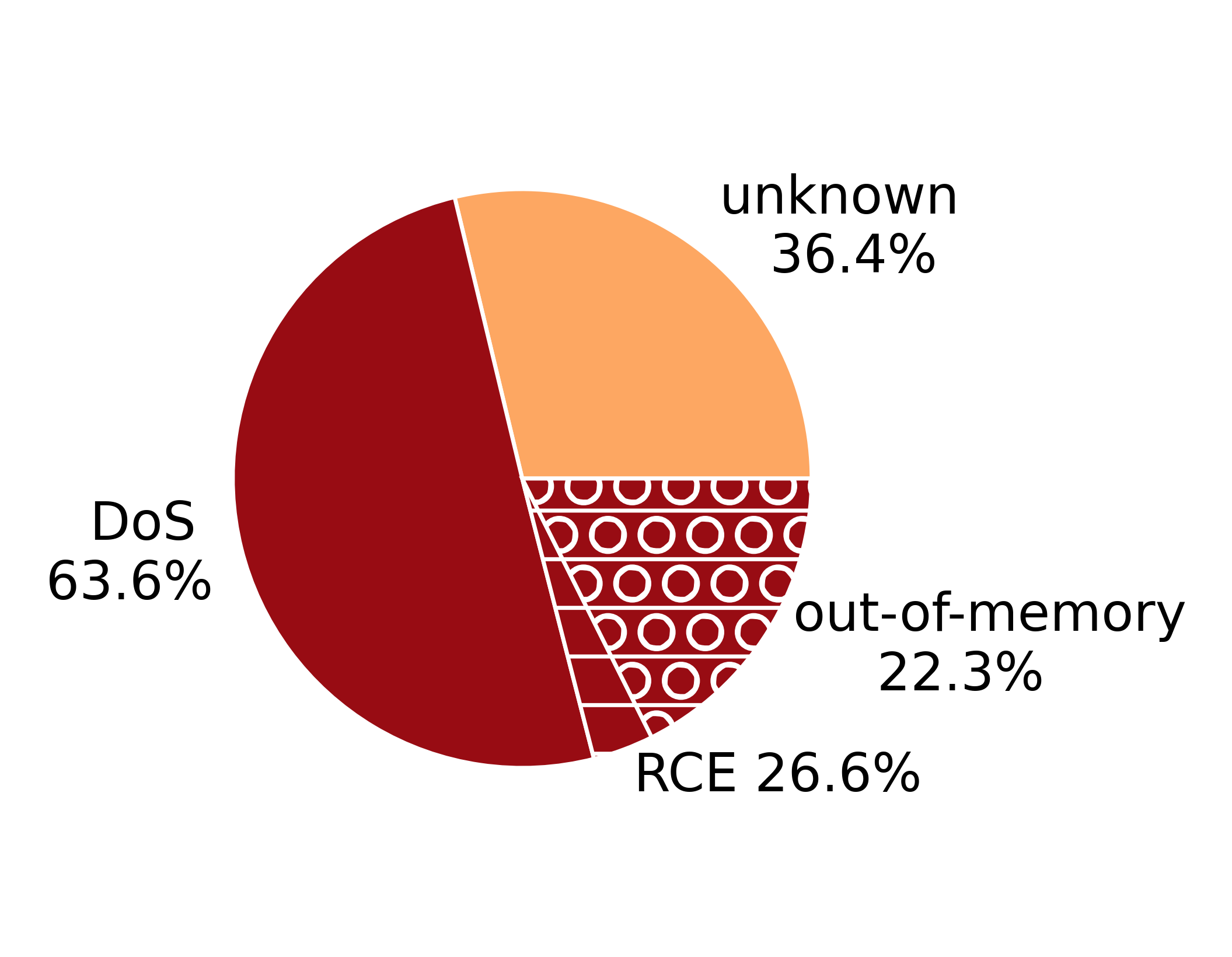}
    \vspace{-30pt}
    \caption{{\footnotesize{OctoRPKI}.}}
    \label{fig:octo-vuln}
\end{subfigure}
\vspace{-5pt}
\caption{Distribution of vulnerabilities in relying parties.}
\label{fig:bugs}
\end{figure}

The ability to remotely disable live RPs is an ongoing issue in RPKI. DoS attacks against RPs directly affect BGP ROV protection. Routers require an uninterrupted connection with RPs to periodically sync their caches. If the connection with RPs drops, routers keep the cache alive for only a short window of time, with default wait times ranging from 600 to 7200 seconds, depending on the router software \cite{cisco:rpki,juniper:rpki,frr:rpki}. \\
\indent Table \ref{tab:rp-cve} is a comprehensive list of RPKI CVEs and RP versions affected. Over the years, a multitude of issues were discovered and disclosed, still, the majority of RP deployments continue to be vulnerable to most of the listed attacks due to lack of software upgrades by operators. Many of the attacks in Table \ref{tab:rp-cve} can be triggered in continuous mode - every time an RP lands on the malicious PP, the crash is triggered anew thus preventing the RP from finishing the validation process. Due to RPs not possessing any mechanisms to bypass misbehaving PPs, they land in a state of perpetual fail-and-restart until operators manually intervene. If an RP is unable to finish a validation cycle without crashing midway, it cannot generate VRPs. The routers connected to them can no longer receive VRPs and after a short wait, they flush their caches and stop enforcing ROV. All these vulnerabilities are easily triggered at no considerable cost from entities that enter the RPKI ecosystem as PPs. Many DoS-inducing errors are caused by malformed data, so they can be triggered either by an attacker or on accident. As a result, the PPs that trigger these issues have plausible deniability on accusations of malicious behavior and there are no protocols to force them to fix this behavior promptly, or exclude them from the RPKI tree. \\
\indent We measure the RP distribution for December 2023 and map the spread of vulnerabilities in the RPKI ecosystem according to Table \ref{tab:rp-cve}. We qualify all RP instances whose agent header version matches with a CVE as vulnerable to the CVE's attack. We do not assume operators independently patched the software. RP software implementations are complex and arbitrary changes from people who are not familiar with the software and protocol can break functionality.\\
\indent Routinator, OctoRPKI and Fort are the only active RPs with known vulnerabilities. In Figure \ref{fig:bugs}, we highlight Routinator and OctoRPKI, two of the most popular RP implementations. Figure \ref{fig:rout-vuln} shows that 37.1\% of Routinator instances are safe from all known vulnerabilities. The remaining 62.8\% are vulnerable to various remotely triggered DoS attacks, 57.9\% of whom are additionally vulnerable to path traversal attacks, and 32.7\% are further affected by RPKI cache poisoning. According to the mappings of Figure \ref{fig:octo-vuln}, 36.4\% of OctoRPKI instances do not have a version in their agent header, and therefore we cannot fully classify their vulnerability status. However, due to missing software upgrades in light of the DoS exploits discovered in \cite{mirdita2023cure}, 100\% of OctoRPKI instances are vulnerable to DoS. To verify this claim, in the absence of feedback from the developers, we further looked at the source code and established that the processing methods that raise most of the errors in \cite{mirdita2023cure} have not been changed since the code's first commit.
26.6\% of those RPs are additionally vulnerable to remote code execution (RCE), and at least 22.3\% are vulnerable to DoS, RCE and out-of-memory attacks. Additionally, only 8.8\% of Fort instances are vulnerable to DoS.
\vspace{-10pt}
\indent Our measurements show that the majority of deployed RPs are vulnerable to easy-to-launch disabling attacks, which can lead to ROV downgrade. With the increased deployment of RPKI in major companies \cite{aws-rpki,twt-rpki}, the interest of adversaries to tamper with ROV validation will only grow, and operators across the world continue to use outdated or vulnerable RPs. 

\subsection{Pervasiveness of Bad Software}

\begin{figure}[h!]
  \centering
  \includegraphics[width=0.4\textwidth]{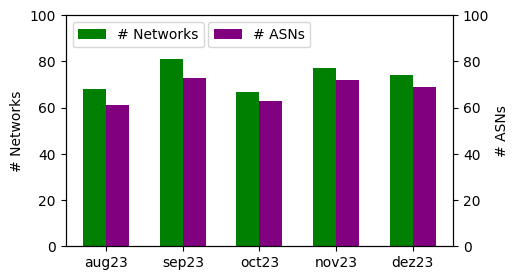}
  \vspace{-10pt}
  \caption{RIPE NCC validator usage trend.}
  \vspace{-10pt}
  \label{fig:ripe-r2}
\end{figure}

We observe three causes that lead to vulnerabilities in RPs: \circled{1} unpatched RPs \circled{2} outdated RPs \circled{3} discontinued RPs. Most existing vulnerabilities in deployed RPs are due to outdated software versions, despite availability of patched upgrades. This issue accounts for 100\% of vulnerable Routinator and Fort instances.
Another issue is unpatched RPs despite known vulnerabilities. This affects 100\% of OctoRPKI instances. We find that a subset of the routing community still uses discontinued software. 
RIPE stopped maintaining its validator in June 2021 and urged users to switch applications \cite{discontinued}. However, as shown in Figure \ref{fig:ripe-r2}, many are still using the discontinued validator in 2023. 
As of December 2023, we map running RIPE validators to 69 ASes.
Due to the severely outdated nature of the software, its continuous usage to inform BGP routing decisions represents a security risk.

\subsection{Attacks for Stalling Relying Parties}

A recurring theme in Table \ref{tab:rp-cve} is stalling attacks. This vulnerability is a complex compound attack that exploits naive RPKI validation processing, while emulating benign network errors. The result is an attack that blinds victim RPs from downloading content from targeted PPs for long periods of time, without causing crashes or anomalous issues. 
Some CVEs already address aspects of this attack. However, proposed solutions only involve basic timers and iteration counters, all of whom fail to actually stop the viability of this attack. \\
\indent 100\% of RP deployments are vulnerable. This attack exploits the inherent processing naivety of the RP. RPs traverse the entire RPKI tree in good faith, and it can neither distinguish between malicious or benign branches, nor prioritize efficient paths. Due to the unconstrained size of the RPKI tree, an attacker can insert an unlimited number of malicious nodes, and therefore an RP can be stuck for many hours on a single validation interval. Stalling exploits the eager validation processing algorithm that does not account for malicious or malfunctioning subtree corner cases. This error can be fixed via a redesign of the validation process that can discriminate between benign and malicious subtrees \cite{hlavacek2023beyond}.

\section{RPKI Ecosystem Longitudinal Analysis}\label{sec:errors}

In this section, we expand existing research of RPKI environments to include our measurements of RPs$\xrightarrow{}$PPs communication trends. This interaction is the linchpin of the RPKI ecosystem. We observe the conditions of PP repositories and their content management efficiency from the standpoint of an RP client. We analyze connections and quantify error rates, their sources and consequences. Additionally, we also measure how errors affect the validation process. \\
\indent We measure the RPKI network with our monitor RP, which periodically fetches the global state of RPKI repositories and logs information about content, connection quality, and errors. This is a longitudinal study that covers two periods: September 2022 - February 2023 (R1) and August 2023 - December 2023 (R2). We use the collected data to extract trends and observations about PP health and RP$\xrightarrow{}$PP connection patterns. 
This study was conducted over two different time periods to observe how these trends evolve as RPKI continues to expand.

\subsection{ROA Content Trends}

\begin{figure*}[t!]
\scriptsize
    \centering
    \includegraphics[width=1.0\linewidth]{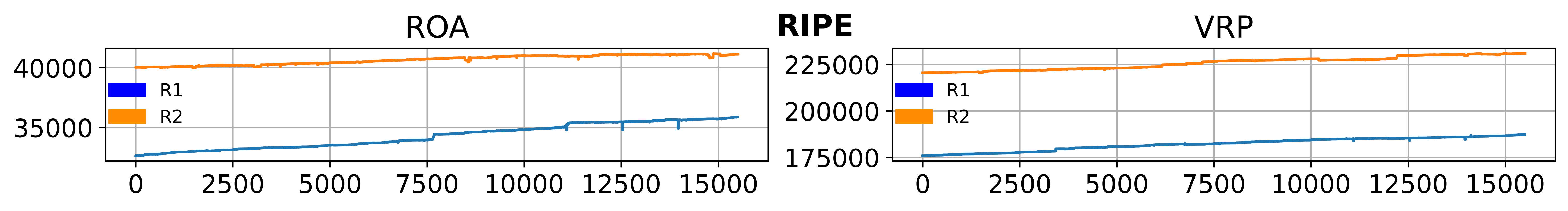}
    \vspace{-5pt}
    \includegraphics[width=1.0\linewidth]{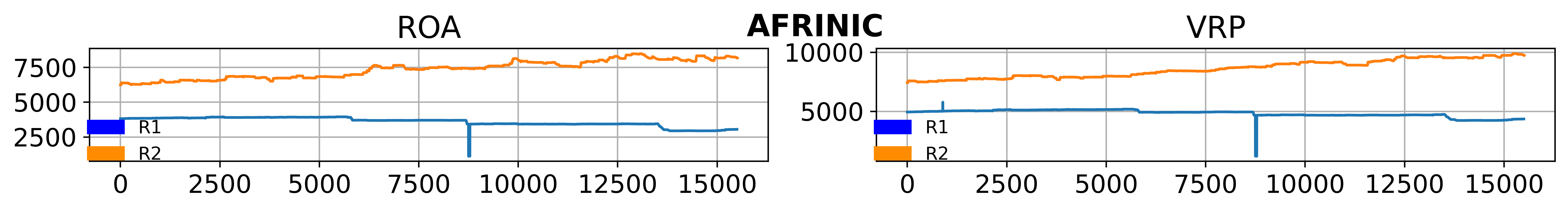}
        \vspace{-5pt}
    \includegraphics[width=1.0\linewidth]{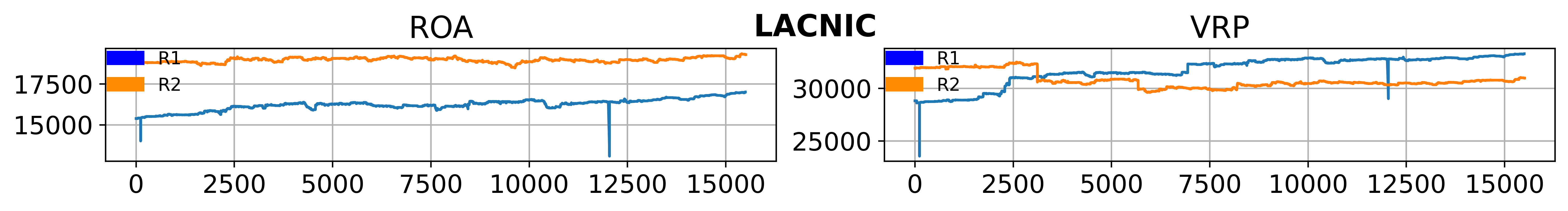}
        \vspace{-5pt}
    \includegraphics[width=1.0\linewidth]{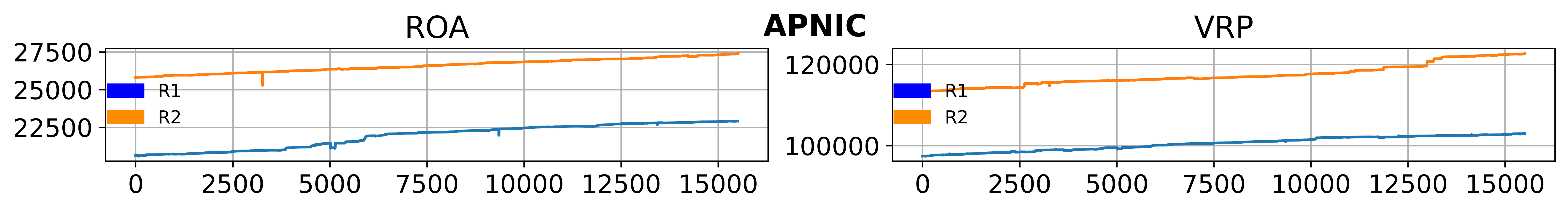}
        \vspace{-5pt}
    \includegraphics[width=1.0\linewidth]{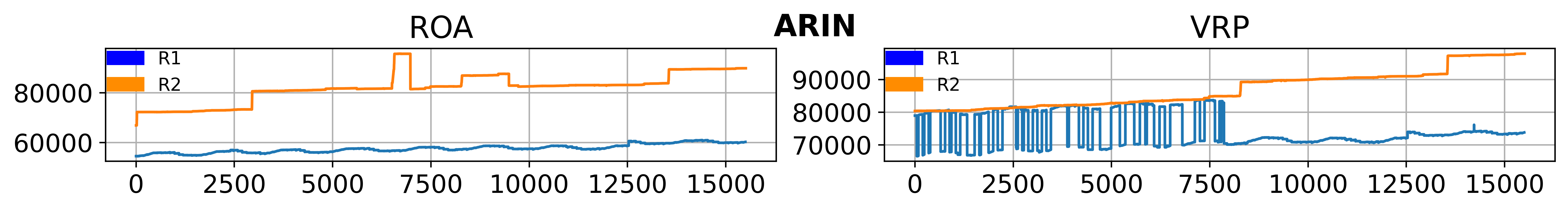}
    \vspace{-10pt}
    \caption{Number of ROA files and VRPs issued during R1 and R2.}
    \label{fig:roa-trend}
\end{figure*}

In this section, we illustrate the correlations of ROAs and VRPs from the perspective of an RP instance. Figure \ref{fig:roa-trend} shows the historic trends of ROAs and VRPs per RIR, over R1 and R2 measurement periods. The values in the graph represent the ROA and VRP content of the RIR's own PP and all delegated PPs attached to that RIR. We differentiate between ROAs and VRPs in two graphs in order to illustrate the correlations between both entities. ROAs can contain multiple VRPs, thus a drop or spike in the total number of ROAs does not necessarily correlate to a change of equal proportions in VRPs, e.g., operators can aggregate existing VRPs into fewer ROA files, or add new VRPs to existing ROAs. \\
\indent Over the two measurement rounds, we observe a gradual increase of ROAs across all RIRs, and a proportional increase of VRPs across most RIRs. This observation is in line with the steady increase of RPKI deployment worldwide. We also observe the presence of frequent blackouts throughout all RIRs. An outage is most dangerous when it lasts at least for multiple hours. Outages of this length lead to routers dropping the missing VRPs due to staleness.\\ 
\indent Data shows that RIPE, APNIC, ARIN and their child repositories are the most stable with the fewest or short-lived blackouts. Most errors characterizing these repositories appear temporarily, only lasting a few validation rounds. The major instability in the ARIN VRP graph during R1 is the result of planned infrastructure tests\cite{arintest}.
However, LACNIC and AFRINIC frequently show instability. These RIRs do not offer easy ways for clients to configure delegated RPKI, so the most resources are centrally managed by the RIR repository. As a result, internal uncaught misconfigurations affect all resource holders. For example, over a 4-day period in early November 2023, $\sim$250 VRPs from AFRINIC were not being correctly propagated due to certificate generation errors, namely {\it certificate is not yet valid} and {\it certificate has expired}. \\
\indent Our measurements show that RPs are not the only RPKI software components that can fail the process. PP object management is just as important and our data suggests that back-end RPKI object generation and maintenance is affected by various issues, including bad certificate maintenance and connection problems. These issues have direct consequences on VRP status caches in routers, thus blackouts and instabilities lead to silent ROV disabling. 

\subsection{Validation Processing Patterns}\label{sec:ref-interval}

\begin{figure}[H]
    \centering
        \includegraphics[width=1.0\linewidth]{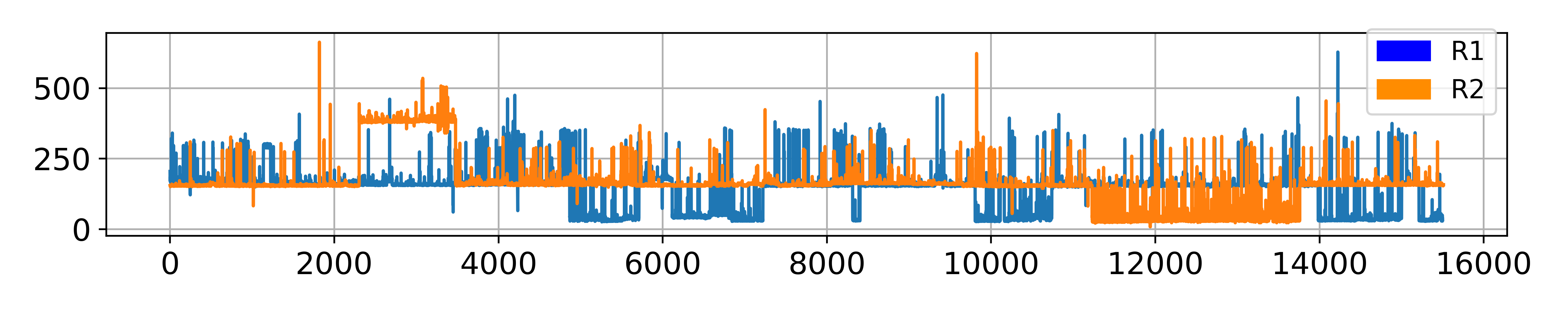}
\vspace{-20pt}
    \caption{Duration of validation intervals for R1 and R2.}
    \label{fig:validation-duration}
\end{figure}

The validation process is the core algorithm which builds the VRP cache for BGP routers. This process is heavily dependent on the integrity and efficiency of RP$\xrightarrow{}$PP communications: their errors and delays. We analyze the RP validation logs and measure the effect of PP behavior and health effects on RPs.\\ 
\indent We observe an increase in average processing time, as expected due to the increase of RPKI deployment. Specifically, during R1 in Figure \ref{fig:validation-duration}, the mean runtime interval is 122 seconds with a minimum of 26 seconds, and a maximum of 628 seconds. We see that the processing times for R1 are highly unstable and oscillate a lot. In R2, we observe two anomalies; the first one is a slowdown that lasted for $\sim$14 days, and the second is a speed-up that lasted for $\sim$15 days. 
During R2, sans anomalous periods, the mean processing time is 161 seconds, minimum is 56 seconds, and maximum is 663 seconds. Figure \ref{fig:validation-duration} clearly shows a stabilization of processing times during R2.\\
\indent Our measurements show an average increase of 39 seconds from R1 to R2. During the transition period, the number of PPs increased from an average of 50 to over 60. We observe that timeout errors cause noticeable slowdowns in the validation interval, even when they happen only to a handful of PPs. Additionally, we observe that same error types do not necessarily cause the same amount of delay. This is due to the nature of the network instability that causes the error in the first place. Connections will either drop due to timeouts, stall for arbitrary times due to low bandwidth, or the webserver resets and drops the connection due to malfunctions. However, our measurements suggest that RPKI is gradually improving its stability and repository resilience.\\
\indent \textit{The 14-day anomaly in R2} was caused by 3 separate but persistent PPs: each costing the validation interval approximately 70, 86 and 127 seconds respectively. The error types in question are \textit{"tcp connect error: Connection timed out"} and \textit{"error trying to connect: Connection reset by peer"}. Within the 14-day anomaly period the average processing time was 363 seconds, hitting maximum processing of 535 seconds. \\
\indent \textit{The 15-day performance uptick in R2} was the result of timeout-error-free validation intervals. Our analysis shows that over this period, in particular during the lowest lasting validation intervals, there were few to no timeout errors and therefore no delays during connection. Upticks in processing time that go back to the most common values are instances of PP connections temporarily under-performing resulting in delays and timeouts. The average processing time during this period is 59 seconds and the median is 32 seconds. \\
\indent Our observations show that the most common processing times during low error validation intervals are in the range 30-40 seconds. Ordinary average validations last at least 5x longer than the absolute necessary processing time due to an amalgam of common networking errors, and the time requirements get at least 10x longer when at least 5\% of PPs malfunction. Full RPKI deployment requires the introduction of additional PP entities in the system, including new processing steps, large amounts of additional data load, and higher wait tolerance for lagging and misbehaving PPs. These discrepancies between necessary processing time and observed processing times, all due to PP misbehavior and network instability, suggests that future full deployment of RPKI as central repository for BGP security auxiliary files, faces adoption bottlenecks due to slow and inefficient processing times.

\begin{table}[H]
\renewcommand{\arraystretch}{0.7}
    \centering
    \begin{tabular}{c|c}
         \textbf{TimeOut Errors} & \textbf{Non-Timeout Errors} \\ \hline \hline
          Connection timed out  & 502 Bad Gateway \\\hline
          Connection refused & 504 Gateway Timeout \\\hline
          Connection reset by peer & malformed XML \\\hline
          Name or service not known & CertExpired \\\hline
          operation timed out  & 404 Not Found \\\hline
          Network is unreachable& - \\\hline
          No route to host  & - \\\hline
        \makecell{rsync error: timeout \\ waiting for daemon connection} & - \\\hline
        rsync error: timeout waiting & - \\\hline
        \makecell{connection closed \\ before message completed} & - \\\hline
    \end{tabular}
    \vspace{-5pt}
    \caption{Types of errors observed.}
    \vspace{-10pt}
    \label{tab:my_labels}
\end{table}

\subsection{Errors in RPKI}

We collect and analyze the errors that occur during the validation process and categorize them in two classes, see Table \ref{tab:my_labels}. Timeout errors are triggered due to timing and networking issues. Non-Timeout errors are caused by data processing issues.
  Most errors in the timeout column are caused by DNS resolution issues, faulty networks, overloaded webservers or low bandwidth that lead to connection slowdowns.

\begin{figure}[H]
    \centering
    \includegraphics[width=1.0\linewidth]{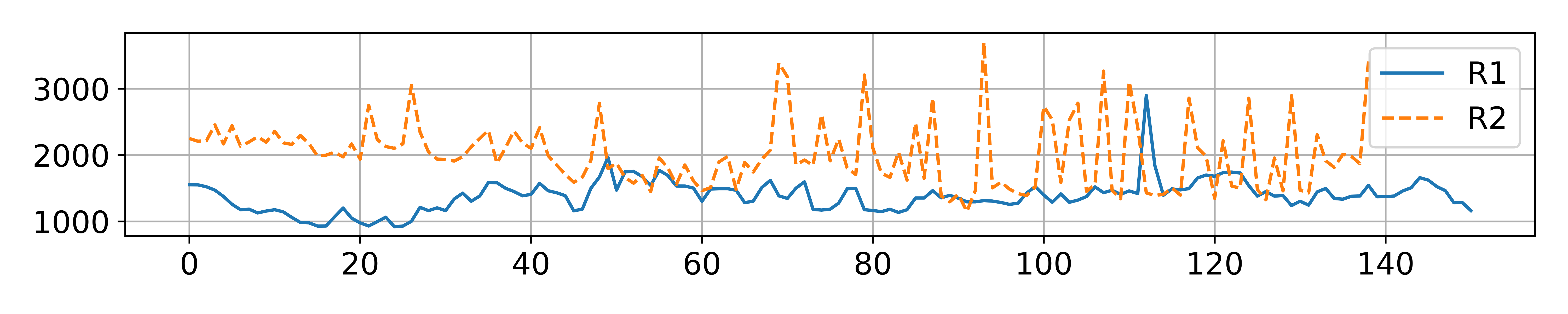}
    \vspace{-20pt}
    \caption{Daily error rate for R1 and R2.}
    \label{fig:daily_error_r1_r2}
\end{figure}

To quantify the effect of errors on validation time, we calculate the Spearman correlation \cite{myers2004spearman} between the rate of occurring errors and processing time per validation round.
Notably, the correlation factor in R1 and R2 for timeout errors is 0.36 and 0.47 respectively, both of which denote moderate correlation.
The lack of an overly strong correlation can be explained based on the nature of the RPKI environment:\\
\indent {\bf Processing time} is influenced by the frequency of changes in the RPKI tree and the data inside repositories. These are fluctuations independent of errors. The more data an RP has to process, the longer it takes. Furthermore, RPKI objects expire within short but irregular intervals, therefore, RPs perform random re-computations due to routine object re-issuance. \\
\indent {\bf Network congestion}, even when they do not lead to errors, can cause slowdowns, disruptions and retries, thus influencing processing time without any logged error entry.\\
\indent {\bf Not all errors are made equal}. The same error type can cause different delay times over different repositories. This happens due to the ephemeral nature of congestion. Delays can last for an arbitrary amount of time, and either break off abruptly or drag the connection until it hits the timeout limit. Errors result is nondeterministic delay intervals.\\
\indent In Figure \ref{fig:daily_error_r1_r2} we show a summarized view of daily errors for R1 and R2. The \textit{x-Axis} is the number of days and the \textit{y-Axis} shows the aggregate daily errors of all types. 
The mean daily error rate increased by 45.2\%, from an average of 1382 in R1 to 2008 in R2. The minimum daily error rate has also gone up from 920 in R1 to 1148 in R2, meaning a 24.7\% increase. This shows that even the rates in the best performing days, are now over 24.7\% higher than less than a year prior. 
We observe the overall share of timeout errors increased from 64.9\% in R1 of all errors to 80.5\% of all errors in R2. This is an increase of 24\% of the timeout error coverage over the two measurement periods. Non-Timeout Errors, on the other hand, show a decrease. In R1 35.1\% of errors were non-timeout errors while during R2 only 19.5\%. This amounts to a 44\% decrease. Timeout errors were the most common error type, and they keep gaining ground compared to processing errors.
These trends suggest that RPKI deployments are improving their RPKI object generation and maintenance, but network sanitization and bandwidth still requires improvement.\\
\indent In terms of error frequency, timeout error rates almost double in R2 compared to R1. rsync timeouts were the most common errors during R1, followed by RRDP connection timeouts and DNS resolution errors. During R2, we see that the primary error type is DNS resolution, followed by rsync timeouts. 
The fact that rsync continuously gets called upon, and still fails to deliver, shows us these are errors leading to total unavailability of PPs. Furthermore, we observe a visible decrease in non-timeout processing errors from R1 to R2. The most persistently common error remains CertExpired which is due to insufficient back-end management of RPKI certificates. According to logs, it is more prevalent in centralized PPs that handle multiple CAs, possibly due to insufficiently fast processing and re-generation of expired objects and certificates. On some occasions, this process is manually performed by a human, as used to be the case for some operators \cite{aws-rpki}. Additionally, lack of adequate webserver management is consistently a major source for connection errors in RPKI, with 404, 502 or 504 error codes being the most common.

\subsection{RP Connectivity Analysis}

During R2, we monitor outgoing DNS requests from our RP instance. Over a 3-month period, we analyze the type and trends of queries an RP sends and responses it receives, per validation round. We let our RP run independently and monitor outgoing requests and incoming responses.\\
\begin{figure}[t!]
    \centering
    \begin{subfigure}{0.5\textwidth}
    \includegraphics[width=\linewidth]{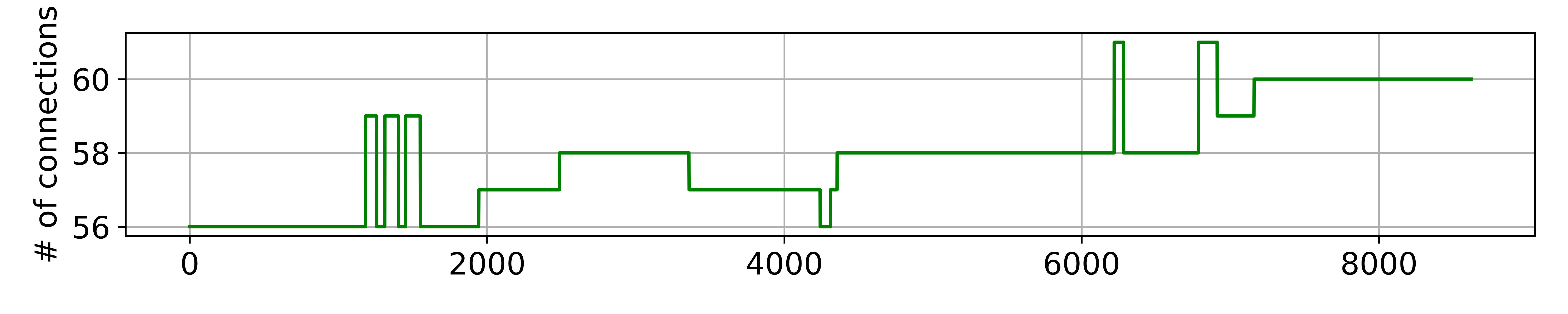}
    \vspace{-15pt}
    \caption{Unique RRDP requests per validation interval.}
    \label{fig:mon-rrdp-pp}
    \end{subfigure}
    \begin{subfigure}{0.5\textwidth}
    \includegraphics[width=\linewidth]{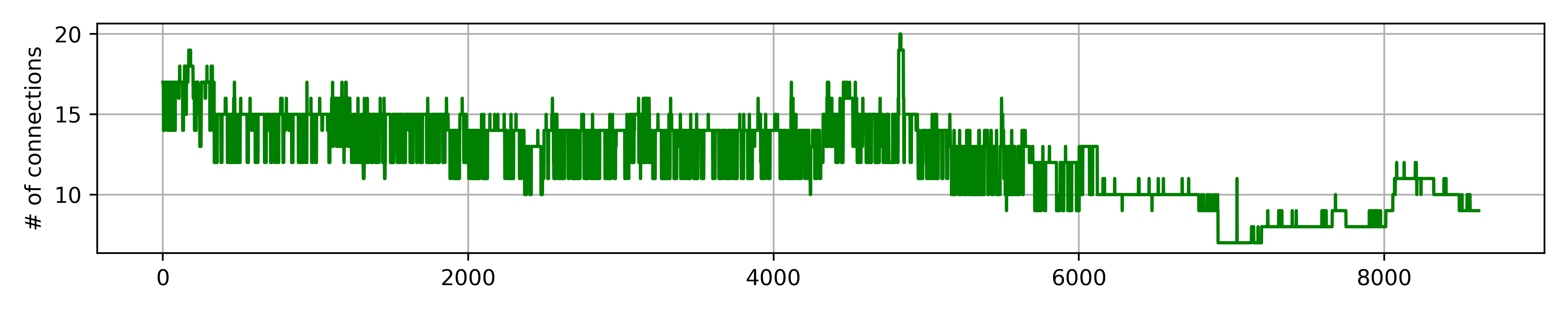}
        \vspace{-15pt}
    \caption{Unique rsync requests per validation interval.}
    \label{fig:mon-rsync}
    \end{subfigure}
     \vspace{-15pt}
    \caption{RP unique requests.}
     \vspace{-10pt}
    \label{fig:dns-res}
\end{figure}
\indent Figure \ref{fig:mon-rrdp-pp} shows the number of unique RRDP requests that an RP sends every validation interval. This graph depicts the evolution of the number of PPs available on the Internet. We can observe the gradual introduction of 7 new PPs, increasing the total from 56 to 63. The fairly stable nature of this graph is due to RRDP being the default fetching protocol of RPKI, so there are few oscillations: only when PPs enter or disappear from the Internet does the number of unique requests change. Occasional oscillations can be explained as the repositories containing the pointers to child PPs malfunction, or delegate certificates are not updated correctly. In Figure \ref{fig:mon-rsync}, we see rsync requests during the monitoring period. The average number of rsync requests is $\sim$10 per validation cycle. Considering that rsync is the backup protocol when RRDP fails, we observe that an average of 17.3\% of RRDP requests fail every validation round, and need to resort to a backup protocol.\\
\begin{figure}[t!]
    \centering
    \begin{subfigure}{0.5\textwidth}
    \includegraphics[width=\linewidth]{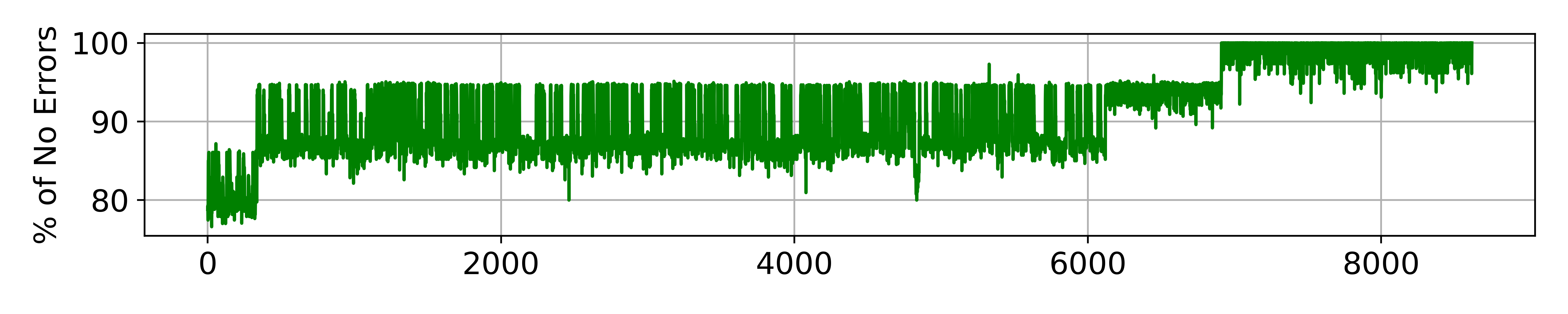}
        \vspace{-15pt}
    \caption{Successful DNS rates.}
    \label{fig:mon-noerror}
    \end{subfigure}
    \begin{subfigure}{0.50\textwidth}
    \includegraphics[width=\linewidth]{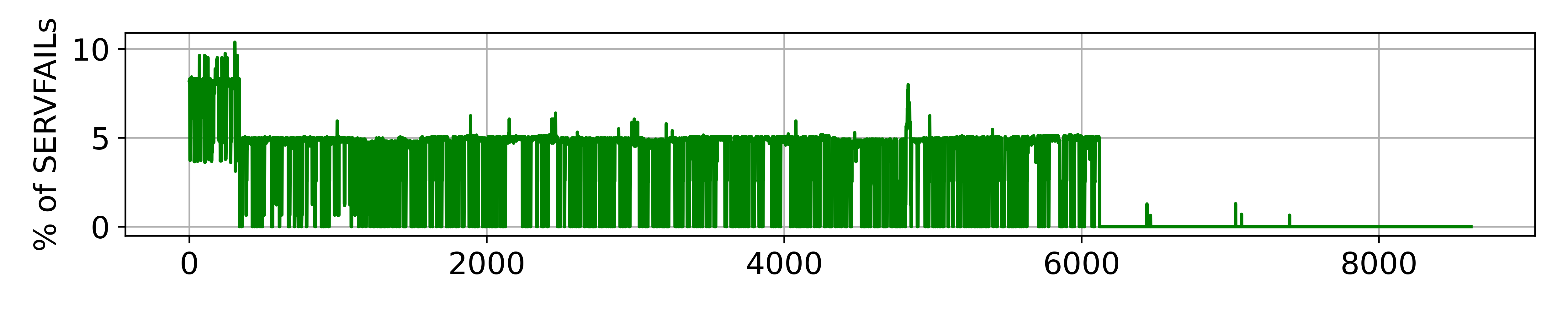}
        \vspace{-15pt}
    \caption{SERVFAIL DNS error rates.}
    \label{fig:mon-servfail}
    \end{subfigure}
    \begin{subfigure}{0.50\textwidth}
    \includegraphics[width=\linewidth]{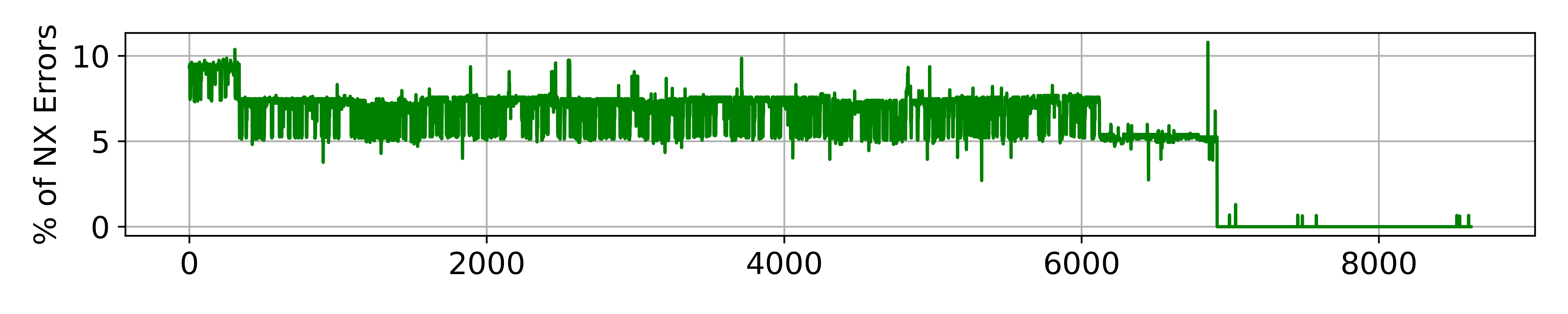}
        \vspace{-15pt}
    \caption{NXDOMAIN DNS error rates.}
        \vspace{-5pt}
    \label{fig:mon-nx}
    \end{subfigure}
    \caption{DNS response types.}
     \vspace{-10pt}
    \label{fig:dns-res}
\end{figure}
\indent We analyze DNS responses, including retries, that our RP receives during validation rounds. Figure \ref{fig:dns-res} shows the percentual rate of DNS responses received for each validation run. Figure \ref{fig:mon-noerror} shows the rate of successful DNS Requests is on average 88.7\% until October. In November, the numbers go up to 97.9\%. During September and October, an average of 3.5\% of DNS Requests returned SERVFAILs, see Figure \ref{fig:mon-servfail}. In November, we observe a network shift where the SERVFAIL error rate plummets to 0.12\%. Further, in Figure \ref{fig:mon-nx} we show another typically occurring DNS error, NXDOMAIN. Prior to the improvements in November almost 6.83\% of connections return a NXDOMAIN response, then the rate goes down to 0.89\%. During this measurement period, we observe an increase in network maturity and stability of PP providers. The improved DNS performance coupled with the marked decrease in non-timeout processing errors suggests that PP operators are focusing in solving network issues and improving the efficiency of their services.

\subsection{Consequences of Errors}

Our measurements highlight a range of issues plaguing RPKI deployments. We observe persistent problems with PP object management. Expired certificates and malformed objects are re-occurring errors. PPs also suffer from connection issues due to DNS errors or misconfigured webservers. These issues lead to major ROA oscillations from the RP standpoint, resulting in downgraded ROV protections despite legitimately issued ROAs. Research done by \cite{mirdita2023cure} supports our observations. The authors discover that even if PPs generate legitimate objects, RPs are prone to discarding them anyway due to non-RFC conform processing idiosyncrasies. \\
\indent The RPKI ecosystem is characterized by subpar network hygiene for many PPs. Networking delays and timeouts are common. Errors are so prevalent that periods with few or no errors appear as graph anomalies. Persistent daily networking errors lead to a validation process at least 5x slower than the baseline processing without delaying errors. When PPs malfunction, processing time increases at least 2.5x times more than the average processing time. The authors in \cite{hlavacek2023beyond} propose full deployment projections based on non-delayed connections. According to these projections, even if conservatively considered, full RPKI deployment may lead up to 41 minutes of download time per validation round, assuming perfect networking conditions and mere delta synchronizations.\\
\indent Our study suggests that routine networking bottlenecks and errors do not allow for error-free connections to become norm. According to network condition trends, the projected full deployment would require 5x more processing time than the baseline. Even if these theoretical projections do not manifest wholly, if we couple full RPKI deployment, increased number of PPs, increased number of CAs and objects, and all BGP security protocols RPKI is slated to cover, with existing rates of network slowdowns, bottlenecks and delays, global RPKI deployment risks becoming slow and inefficient.

\section{The Human Factor}\label{sec:eval}

We conduct new measurements on previously untapped facets of RPKI. 
Our analysis establishes from multiple vantage points, various persistent security issues caused primarily by RPKI stakeholders: developers and operators. We show that subpar deployment practices and lax security-awareness are major issues for global RPKI deployments. \\
\indent {\bf Publication point operators.} PP service quality is important for the correct propagation of RPKI data. When a PP is misconfigured, all RPs are affected by delays. Our measurements show that the most commonly triggered errors by PPs are timeouts and DNS issues. Processing time increased by 31\% in less than a year, while still halfway to full RPKI coverage of Internet resources, and just starting to support additional BGP security algorithms, all of whom require more content and processing logic. Measurements in Figure \ref{fig:daily_error_r1_r2} show that errors in RPKI have risen. PP operators need incentives to mind to their security when entering the ecosystem. These incentives can be enforced by monitoring authorities, software alerts, and improved algorithms that discriminate between good and bad RPKI branches.\\
\indent {\bf Relying party developers.} RP behavior and output influences global inter-domain routing. Therefore, it is important for software developers to address issues in a timely manner, implement software according to protocol requirements, and be mindful of network sanity checks. Recent research \cite{mirdita2023cure} documents various issues with RP development. First, some implementations do not follow RFC requirements, thus instances of processing disparities were discovered. This processing discrepancy has widespread practical consequences. Resource owners who issue valid ROAs can suffer a silent downgrade of their resource protection in regions of the Internet using a badly designed RP. Researchers discovered over 8000 network prefixes whose protection was downgraded due to one particular RP processing RFC guidelines more strictly than others. The certificates of the downgraded resources were valid and benign, they just contained a superfluous certificate header not included in the standard. 
This was enough for an overeager RP to flush the certificates and accompanying resources. We also notice that many DoS-triggering issues that researchers discovered in OctoRPKI, have not yet been patched by developers. 
OctoRPKI is the third largest RP deployement on the Internet and all its instances are vulnerable to remote shut downs. Developers' indifference on the security of their RPKI software is an issue that can slow down full RPKI adoption and should be discouraged. \\
\indent {\bf Relying party operators.} RPs are the middleman between ROAs and BGP routers. 
The growing use of RPKI globally requires conscientious deployment of RPs. If an RP is vulnerable, the integrity of the data getting fed to a router is not guaranteed. Our measurements show that most RP operators are not working in a security-aware manner. At least 56\% of global RPs are currently vulnerable to at least one known attack, many versions are vulnerable to multiple attacks simultaneously, and 100\% of RPs are vulnerable to stalling attacks. 62.8\% of Routinator instances on the Internet are vulnerable to at least 1 attack vector that can shut down the RP instance completely in a perpetual fail-and-restart mode.
This effectively downgrades ROV protection for the ASes depending on those routers, and for the clients who have issued legitimate valid ROAs. The same applies to 100\% of OctoRPKI instances. Most of the aforementioned issues have already been addressed in public patches, but the uptake is slow. Operators are simply not upgrading their software fast enough. A surprising 2.1\% of global RPs are still using RIPE NCC RPKI Validator, software that was discontinued in 2021. Furthermore, all global RP implementations are vulnerable to stalling attacks. Given this constellation of issues, it is clear many RP operators suffer from lax security-awareness.

\section{Future Work}\label{fw}

\indent {\bf Overhaul the protocol.} RPKI's primary purpose is to protect BGP routing from hijacks. However, its own design and security parameters leave it vulnerable to external attacks and benign failures. Resource delegation in RPKI is too flexible. CAs can infinitely delegate the same resources, thus enabling the creation of infinite CA chains. Infinite delegations coupled with low-volume out-of-band packet loss enable large-scale stalling attacks. Individual vendors have introduced solution proposals, but they were mere patches containing arbitrary thresholds, that did not solve the problems.
The validation algorithm treats all PPs in good faith, but PPs can be adversarial. A naive approach that trusts all repositories with no mechanisms to avoid malicious behavior, can be exploited to downgrade RPKI protection. 
Existing projections of full RPKI deployment scenarios highlight the vulnerability of the system from misbehaving PPs. It is important for RPKI algorithms to incorporate self-preservation. 
Future work requires operators to introduce intelligent algorithms and decouple VRP generation from RTR communication so that anomalous validation intervals do not interfere with the RTR protocol.\\
\indent Additionally, RP implementations often exhibit unequal behavior due to vague RFC directives and overeager parsers, depending on where one stands on the debate between stricter RFCs or laxer processing. These processing inconsistencies directly affect BGP routing decisions and silently downgrade protection for arbitrary users without their knowledge - or the knowledge of the routing operators themselves. The RPKI standards demand more work from IETF and IRTF contributors to improve clarity and avoid arbitrary misinterpretations. \\
\indent {\bf Alternative infrastructures.} 
RPKI is vulnerable to censorship and global blackouts as a result of the hierarchical design of the repository infrastructure, unilateral resource delegation rights and centralized data storing infrastructure, which introduces single points of failure. 
RIRs, even in their current role as the root of trust in RPKI, are neither above legal coercion by local authorities to block resources \cite{ripe-court}, nor self-initiated coercive measures towards clients \cite{afrinic-court}. 
Research has already shown mechanisms for parent nodes to revoke ROAs for child or grand-child delegations with minimal collateral damage. Full parental control of all delegated resources introduces the potential for the manipulation of repository content, thus raising questions on global RPKI object integrity. 
The decentralization of RPKI object storage, re-issuance and revokation, is an important path forward for future RPKI research, as it addresses a structural issue in the RPKI that has critical security and integrity consequences for full RPKI deployment.\\ 
\indent Many works proposed new RPKI designs to target the aforementioned security concerns, however, a major problem with most existing solutions, is that they fall short on compatibility with existing infrastructure. An important aspect of developing alternative solutions, is ensuring backwards compatibility of new solutions with existing deployments. Operators will always be slow to uptake; it took over a decade for RPKI to reach current deployment levels which are promising though still insufficient. RPKI upgrades and improvements must be compatible with existing RPKI deployments, in order to convince old and new operators to uptake the changes. Otherwise, we risk ending up either with purely theoretical solutions, or a balkanized tableau of incompatible RPKI deployments. More work is necessary to improve the RPKI ecosystem without breaking compatibility with the current infrastructure.\\
\indent {\bf Improve deployment via outreach.} Recent RPKI ROV measurements suggest 27\% of ASes fully enforce ROV. While this is a considerable improvement compared to prior measurements, there is still need for more operators to join the effort. RIRs, IETF events and scientific conferences need to invest more resources and effort for operators to have easy access to comprehensive information on RPKI specifics, deployment strategies, benefits and downsides. Surveys \cite{qinunderstanding} show operators are often reluctant to deploy ROV because of errors in software, lack of comprehensive tutorials, scarce documentation for specific requirements and network topology, and low ROA coverage coupled with bad ROA configurations which lead to false positives and the flushing of legitimate BGP announcements. Many of these issues can be addressed by distributing and providing more quality information from the RPKI community at large, both at the level of researchers and Internet authorities such as RIRs, IETF and IRTF.

\vspace{-10pt}
\section{Conclusions}\label{sec:conclusions}

RPKI is the future of BGP security. Major companies and Internet backbones worldwide are deploying the RPKI protocol to protect their resources from BGP hijack attacks. Despite limited deployment, RPKI has already demonstrably delivered results for all ASes that enforce ROV \cite{twt-rpki}, meanwhile the rest had to struggle with the aftermath of attacks. 
On the other hand, with wider adoption, any failure and vulnerability in RPKI, will affect larger parts of the Internet.\\
\indent RPKI is rife with issues. The RPKI infrastructure is centralized and hierarchical. It allows for non-consensual resource revocations. Further, if one node goes down, everything downstream goes down with it. RPKI trees have unlimited branching capability and RP validators have to graciously visit every node. RPKI is easy to stall thus selectively blinding victim RPs. This results in a silent but devastating downgrade of ROV protection. RPs show processing inconsistencies and contain many of crash-inducing bugs. PPs show lack of rigorous object and server management, which leads to unpredictable slowdowns during processing. To complete the trifecta of issues, there is also the human component. Many operators hesitate to upgrade their software. These issues become even more pressing for RPKI due to the open nature of the environment, where an adversary can monitor and map at low-cost the location of its victim RP or PP, and launch targeted attacks given a known catalog of remotely-triggered errors.\\
\indent In this work, we provide the first mapping of live vulnerabilities in the globally deployed RPKI ecosystem. We quantify the number of RPs that are vulnerable to known attacks and systematize the errors that are most commonly occurring in PPs. Our measurements illustrate that the biggest risk factor to the RPKI ecosystem is human negligence. Our second contribution is a systematization of RPKI security literature with a focus on errors and vulnerabilities, which we quantify to show the actual presence and impact in the live RPKI network. We highlight the promising global acceptance rate of this protocol and systemize the issues that put full deployment at risk. \\
\indent According to our analysis, there is a set of steps to improve and prepare RPKI for full global deployment: \circled{1} Improve the validation algorithm to be able to discriminate between malicious nodes and benign ones. \circled{2} More research on alternative infrastructure implementations utilizing distributed databases and MPC. 
\circled{3} More effort in targeted outreach from authorities such as RIRs, IETF or scientific conferences towards routing operators to foster regular interactions between RPKI developers and operators regarding the latest security risks and best deployment practices.  \\
\indent RPKI is a promising protocol on its way to becoming the most successful BGP security mechanism. 
However, there is still ample work to do to prepare RPKI for full global deployment and fortunately, we see the community moving, albeit slowly, towards this goal.

\section*{Acknowledgements}
We thank Niklas Vogel for his comments on our manuscript. This work has been co-funded by the German Federal Ministry of Education and Research and the Hessian State Ministry for Higher Education, Research and Arts within their joint support of the National Research Center for Applied Cybersecurity ATHENE and by the Deutsche Forschungsgemeinschaft (DFG, German Research Foundation) SFB~1119.

\balance

\bibliographystyle{plain}
\bibliography{ref,bib,sec,main}

\end{document}